\documentclass[useAMS,usenatbib]{mn2e}

\usepackage{graphicx}
\usepackage{dcolumn}
\usepackage{bm}
\usepackage{amssymb, amsmath}
\usepackage{color}

\usepackage{hyperref}
\newcommand \bell     {{\boldsymbol{\ell}}}

\usepackage{aas_macros}
\usepackage{mn2e-bst}

\begin{document}

\title{Background Subtraction Uncertainty from Submillimetre to Millimetre Wavelengths}
\author[Simone Ferraro and Brandon Hensley]{Simone Ferraro$^1$\thanks{E-mail:
sferraro@princeton.edu}, and Brandon Hensley$^1$\thanks{E-mail:
bhensley@princeton.edu}\\
$^{1}$Department of Astrophysical Sciences, Princeton
  University, 4 Ivy Lane, Princeton, NJ 08544, USA
}

\date{\today}
             
\pagerange{\pageref{firstpage}--\pageref{lastpage}} \pubyear{2014}

\maketitle

\label{firstpage}

\begin{abstract}
Photometric observations of galaxies at submillimetre to millimetre wavelengths (50 - 1000 GHz) are susceptible to spatial variations in both the background CMB temperature and CIB emission that can be comparable to the flux from the target galaxy. We quantify the residual uncertainty when background emission inside a circular aperture is estimated by the mean flux in a surrounding annular region, assumed to have no contribution from the source of interest. We present simple formulae to calculate this uncertainty as a function of wavelength and aperture size. Drawing on examples from the literature, we illustrate the use of our formalism in practice and highlight cases in which uncertainty in the background subtraction needs to be considered in the error analysis. We make the code used to calculate the uncertainties publicly available on the web.
\end{abstract}


\section{Introduction}
Emission from galaxies at submillimetre to millimetre wavelengths provides a window into their dust and star formation properties and has thus been studied by numerous ground-based (e.g. SCUBA, SABOCA, LABOCA), balloon (e.g. TopHat, PRONAOS, BLAST), and space-based (e.g. {\it Herschel}, {\it Planck}) missions.

In determining the emission of a galaxy at these wavelengths, careful subtraction of emission from Galactic foregrounds, the Cosmic Infrared Background (CIB), and the Cosmic Microwave Background (CMB) must be performed. Galactic foregrounds should vary smoothly across the sky and are often accounted for by subtracting emission from a region surrounding the target. This technique also removes the mean emission from both the CIB and CMB. However, residual errors due to spatial variations in the CMB temperature and CIB emission remain and in some cases are significant relative to the flux of the target galaxy. Notably for the Small Magellanic Cloud (SMC), CMB temperature fluctuations have provided a partial explanation for submillimetre brightness in excess of predictions by current dust models \citep{Bot+etal_2010, Planck_SMC_2011}, underscoring the need to quantify the background subtraction uncertainty.

In this work, we present calculations of the uncertainty in CMB and CIB subtraction when the surface brightness in an annulus surrounding a target galaxy is subtracted from the target disk. We derive simple formulae to quantify the uncertainty as a function of frequency and aperture size that allow for reliable error estimation without the need for implementing more sophisticated algorithms.

We organise the paper as follows: in Sections 2 and 3 we review the properties of the CMB and CIB, respectively; we discuss other sources of contamination in Section 4; in Section 5 we precisely define the background subtraction procedure and geometry; we present the temperature estimation uncertainty as a function of aperture size and frequency for two geometries in Section 6; we demonstrate our method on objects from the literature in Section 7; and we discuss the scope of our methods in Section 8. We provide a detailed derivation of our results in Appendices A-E. A code for calculating uncertainties in common cases is made publicly available on the web\footnote{http://www.astro.princeton.edu/$\sim$bhensley/software/fluctuations}.

\section{CMB basics}
The CMB is a thermal relic from the hot primordial phase of the Universe. Its spectrum is that of a black body with temperature $T_{CMB} = 2.725$ K. On the top of this uniform background, there are small temperature fluctuations from one point on the sky to another. If we look in the direction $\btheta$ on the sky, we write $T(\btheta) = T_{CMB} + \Delta T(\btheta)$. The fluctuations are remarkably Gaussian, with zero mean and standard deviation of about 110 $\mu$K. These fluctuations are correlated between different points in real space, but decouple in harmonic space (i.e. after spherical harmonic transform), because of translational and rotational invariance. 

Because we are going to restrict ourselves to fairly small patches of the sky (smaller that 10 deg), we will use the flat sky approximation, in which we use Fourier transforms instead of spherical harmonic transforms.
We can then decompose the temperature fluctuations as

\begin{equation}
\ \Delta T(\btheta) = \int \frac{\mathrm{d}^2\ell}{(2 \pi)^2}
e^{i \bell \cdot \btheta} a(\bell)
\end{equation}
Where the $a(\bell)$ are now uncorrelated Gaussian random variables, whose statistical properties are completely determined by their variance, usually denoted by $C_\ell$:
\begin{equation}
\ \langle a(\bell) a(\bell') \rangle = (2 \pi)^2 \ C_\ell
\ \delta_D (\bell + \bell')
\end{equation}
 
The value of $C_\ell$ have been measured with high precision by CMB satellite experiments such as WMAP, Planck, and small-scale, ground based experiments such as ACT and SPT. In this paper, we will use theoretical $C_\ell$ generated with the publicly available software CAMB\footnote{\url{http://camb.info/}}, with cosmological parameters from the Planck 2013 release \citep{2013arXiv1303.5076P}.

\section{The CIB}
The CIB is the roughly isotropic diffuse infrared emission from unresolved dusty star-forming galaxies at high redshift (typically $z \sim $ 1 - 5). Following \citet{2012ApJ...752..120A, 2013JCAP...07..025D, 2012ApJ...755...70R, 2011A&A...536A..18P, 2013arXiv1309.0382P}, we model the CIB emission as the sum of two components, one Poisson component that is dominant at higher $\ell$ and a clustered component that takes into account that these galaxies are biased tracers of the underlying dark matter density.

The angular power-spectrum of the the Poisson and clustered components is well-fit by the following \citep{2013JCAP...07..025D}:
\begin{align}
\frac{\ell^2 C_{\ell}^{CIB, P}}{2 \pi} &= a_p \left( \frac{\ell}{\ell_0} \right)^2 \left[ \frac{\mu^2(\nu, \beta)}{\mu^2(\nu_0, \beta)} \right] \ \mu {\rm K}^2 \\
\frac{\ell^2 C_{\ell}^{CIB, C}}{2 \pi} &= a_c \left( \frac{\ell}{\ell_0} \right)^{2-n} \left[ \frac{\mu^2(\nu, \beta)}{\mu^2(\nu_0, \beta)} \right] \ \mu {\rm K}^2
\end{align}
The value of all of the parameters is reported in Appendix \ref{sec:CIBparams}. The frequency dependence $\mu(\nu, \beta)$ is given by a modified black-body:
\begin{equation}
\mu(\nu, \beta) = \nu^\beta B_\nu(T_d) / G(\nu, T_{CMB})
\end{equation}
Here $B_\nu$ is the black-body function, $T_d$ is the effective dust temperature and $G(\nu, T_{CMB}) = (\partial B_\nu(T) / \partial T )_{T_{CMB}}$ converts between temperature and intensity units, as explained in Section \ref{sec:freq}. This frequency dependence is a good fit for the range 100 GHz - 1000 GHz as shown in \citep{2012ApJ...752..120A}, but should extend to lower frequencies as well.

One comment on units is in order here: For CIB or Radio galaxies, the temperature units \textit{do not} correspond to the brightness or antenna temperature of the sources. They instead correspond to the temperature deviation of a perfect black body from $T_{CMB}$ needed to create the corresponding change in intensity at a particular frequency. To convert them to the more physical intensity units, one must use the function $G(\nu, T_{CMB})$, as explained in Section \ref{sec:freq}.

Note that the CMB fluctuations are suppressed on small scales by Silk damping, therefore the CIB is always dominant in this regime (corresponding to small $\theta_d$).

In the remaining of the paper, $C_\ell$ will refer to the total fluctuation (CMB + CIB). Since they are uncorrelated, their variances (power spectra) add:
\begin{equation}
C_\ell = C_\ell^{CMB} + C_{\ell}^{CIB, P} + C_{\ell}^{CIB, C}
\end{equation}

\section{Other Backgrounds and galactic foregrounds}
In this work we focus on the frequency range 50 GHz - 1000 GHz, where CMB and CIB are the dominant background components.  Towards the low frequency end ($\nu \lesssim $ 150 GHz), the contribution from Radio Point sources (essentially synchrotron emission from background galaxies) becomes comparable in size to the CIB and CMB, and we model it as described in \citep{2013JCAP...07..025D}:
\begin{equation}
\frac{\ell^2 C_{\ell}^{rad}}{2 \pi} = a_s \left( \frac{\ell}{\ell_0} \right)^2 \left( \frac{\nu^2}{\nu_0^2} \right)^{\alpha_s} \left[ \frac{G^2(\nu_0, T_{CMB})}{G^2(\nu, T_{CMB})} \right] \ \mu {\rm K}^2
\end{equation}
Where we take $\alpha_s = -0.5$, $a_s = 2.9$ but note that the amplitude depends on the intensity cut adopted to mask bright point sources. The values of $\nu_0$ and $\ell_0$ are the same as for the CIB and as usual we add $C_{\ell}^{rad}$ to the $C_\ell$ from other sources.

Other sources of background fluctuations such as Thermal (tSZ) and Kinematic (kSZ) Sunyaev-Zel'dovich effects are subdominant in the frequency range considered.

Galactic foregrounds are spatially highly anisotropic and therefore not well suited to be included in our treatment. They can however be important and need to be characterised in the proximity of the direction of observation (for example by the use of dust, HI, and galactic synchrotron maps).

\section{The setup}
In this section we make the problem of subtraction more precise and compute the uncertainty on the residual fluctuations after subtraction of the average temperature in an annulus around our galaxy. We defer detailed derivations to the Appendix and report here the results of practical use.

Suppose that the galaxy in question is fully contained in a circular disk of radius $\theta_d$ and that the average background fluctuation (due to CMB, CIB and radio point sources) in the disk is $\overline{\Delta T_d}$.

\begin{figure}
\centering
    \scalebox{1.0}{\includegraphics{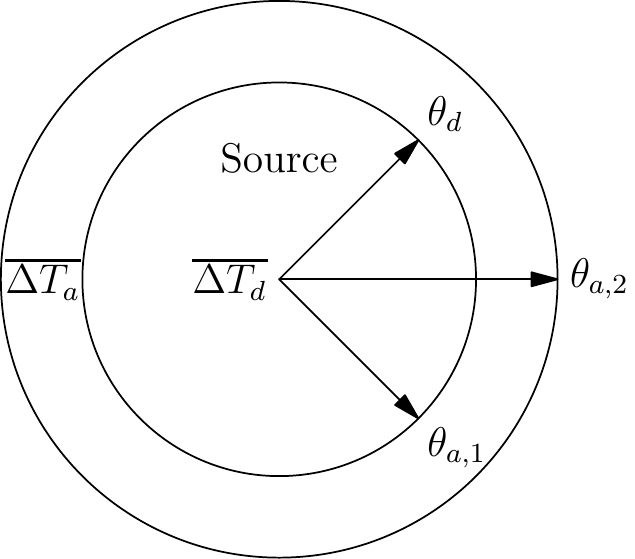}}
\caption{The assumed geometry of the calculation. A galaxy is imaged with a photometric aperture of radius $\theta_d$ and a background subtraction is done with an annular region of inner radius $\theta_{a,1}$ and outer radius $\theta_{a,2}$. The unknown mean temperature fluctuation in the aperture $\overline{\Delta T_d}$ is estimated using the mean temperature fluctuation in the annulus $\overline{\Delta T_a}$.}
\label{fig:geom}
\end{figure}

Since $\overline{\Delta T_d}$ is not directly measurable (the total emission in the disk is often dominated by the galaxy), we also measure the average fluctuation $\overline{\Delta T_a}$ in an annulus around it, contained between radii $\theta_{a, 1}$ and $\theta_{a, 2}$. We will assume that the annulus is not contaminated by the galaxy emission, so that 
we can estimate average temperature fluctuation behind the galaxy as $\overline{\Delta T_a}$ and subtract this value. We ask what is the uncertainty when estimating $\overline{\Delta T_d}$ as $\overline{\Delta T_a}$, i.e. the variance of 
\begin{equation}
\mathcal{D}_{d,a} = \overline{\Delta T_d} - \overline{\Delta T_a}
~~~.
\end{equation}

\section{Results}
We first notice that our guess $\overline{\Delta T_d} = \overline{\Delta T_a}$ is statistically \textit{unbiased}, in the sense that $\langle \mathcal{D}_{d,a} \rangle  = 0$, where $\langle \cdot \rangle$ denotes ensemble average over realisations of the initial conditions, or alternatively spatial average over many galaxies.

Regarding the variance, in Appendix \ref{app:diff}, we show that 
\begin{equation}
\langle \mathcal{D}^2_{d,a} \rangle =
\langle \left(\overline{\Delta T_d}\right)^2 \rangle + \langle \left(\overline{\Delta
  T_a}\right)^2 \rangle - 2 \langle \overline{\Delta T_d} \, \overline{\Delta T_a}
\rangle
~~~,
\end{equation}
where
\begin{align}
\langle \left(\overline{\Delta T}_i\right)^2 \rangle & =  \int \frac{\mathrm{d}\ell}{2 \pi} \; \ell \ b_\ell^2 C_\ell  | W_i (\ell) |^2 \\
\langle \overline{\Delta T_d} \, \overline{\Delta T_a} \rangle &= \int \frac{\mathrm{d}\ell}{2\pi} \; \ell \ b_\ell^2 C_\ell W_a (\ell) W_d(\ell)
~~~.
\end{align}
Here $C_\ell$ are the CMB + CIB multipole coefficients and $W(\ell)$ is the Fourier transform of the window function (disk or annulus). The effect of the finite beam size is encoded in the coefficients $b_{\ell}$, which are the Fourier transform of the beam function, as explained in Appendix \ref{sec:beam}

The formulae above are only valid for circularly symmetric window functions, and the more general case is treated in the Appendix. As an example, we will show results for two common cases of disk and annulus having the same area or the same width.

\subsection{Conversion to Intensity Units}
\label{sec:freq}
\begin{figure}
\centerline{
   \includegraphics[width=8cm]{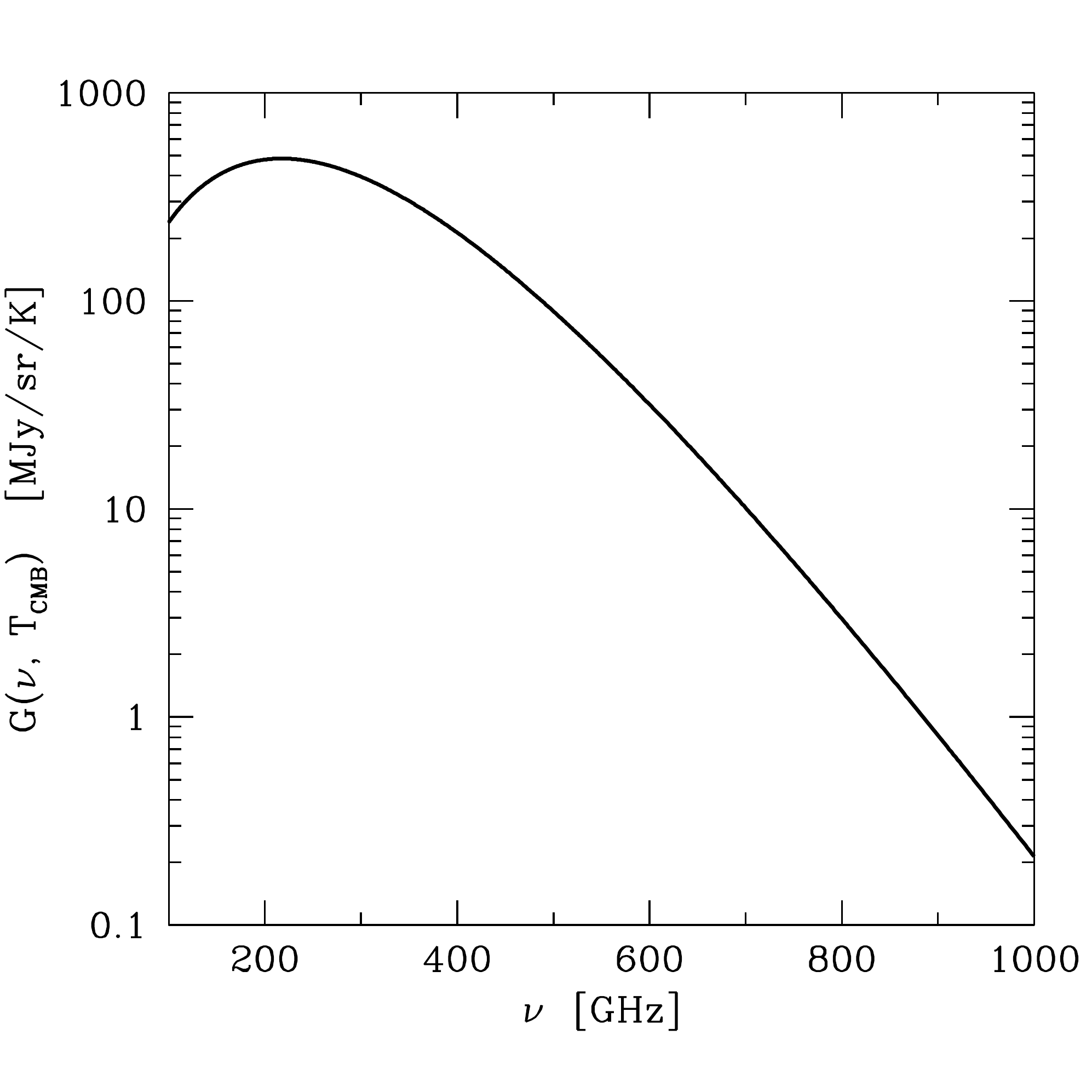}
}
\caption{Conversion factor such that $\Delta I_{\nu} = G(\nu, T_{CMB}) \Delta T$}
\label{fig:K}
\end{figure}

For a uniform black body at temperature $T_{CMB}$ = 2.725 K, the corresponding (specific) intensity is given by
\begin{equation}
I_{\nu}^{(0)} = B_\nu(T_{CMB}) = \frac{2 h \nu^3}{c^2} \frac{1}{e^x -1}
~~~,
\end{equation}
where $x = h \nu / k_B T_{CMB}$.
The fluctuation in intensity due to a temperature fluctuation $\Delta T$ is (to first order):
\begin{align}
\Delta I_{\nu} &= \left( \frac{\partial B_{\nu}}{\partial T} \right)_{T_{CMB}} \Delta T = \frac{2 h \nu^3}{c^2} \frac{x e^x}{(e^x-1)^2} \frac{\Delta T}{T_{CMB}} \nonumber \\ 
 & \equiv G(\nu, T_{CMB}) \ \Delta T
 ~~~,
\end{align}
where we have defined $G(\nu, T_{CMB})$ as the conversion factor between temperature and intensity fluctuations.
It follows that 
\begin{equation}
\langle (\Delta I_{\nu})^2 \rangle^{1/2} = G(\nu, T_{CMB}) \langle \mathcal{D}^2_{d,a} \rangle^{1/2} \ \ .
\end{equation}
$G(\nu, T_{CMB})$ as a function of frequency is illustrated in Figure~\ref{fig:K}.

\subsection{Equal Area Case}
Here we take
\begin{align}
{\rm Disk:} &\ {\rm Radius = \ } \theta_d \\
{\rm Annulus:} &\ {\rm Between \ } \theta_{a,1} = \theta_d {\ \rm and \ } \theta_{a,2} = \sqrt{2} \theta_d
\end{align}
Moreover we take the limit of an infinitely narrow beam, and set $b_\ell = 1$. The inclusion of a finite beam is straightforward and discussed in Appendix \ref{sec:beam} . Explicit expressions for the window function can be found in Appendix \ref{sec:win}.

\begin{figure}
\centering
    \includegraphics[width=8cm]{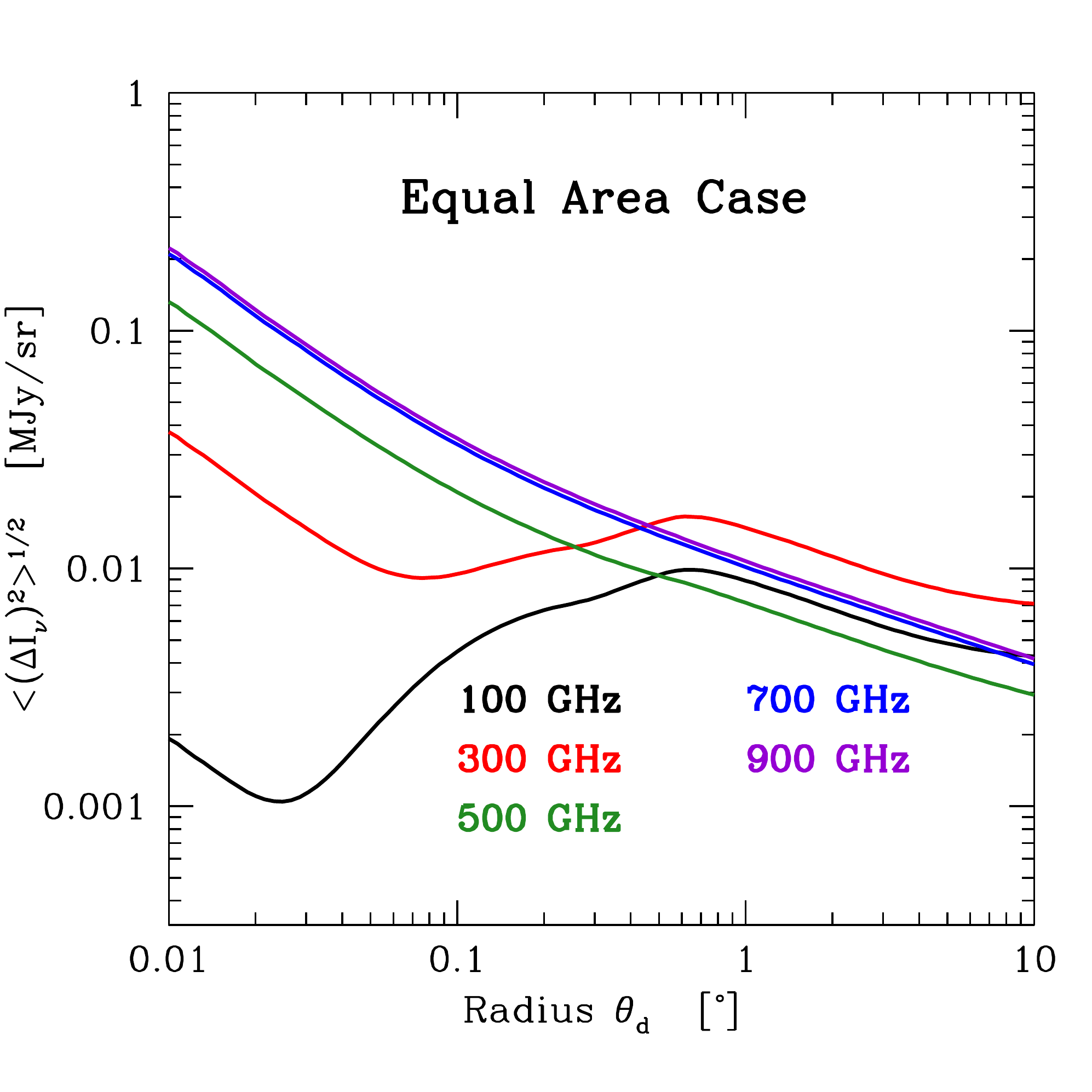}
\caption{Fluctuations in intensity for the equal area case in the limit of an infinitely narrow beam ($b_\ell = 1$). The different curves show fluctuations at 100, 300, 500, 700 and 900 GHz.}
\label{fig:EA}
\end{figure}

In Figure \ref{fig:EA} we show the total (CMB + CIB + Radio sources) fluctuations between the disk and annulus for a few different frequencies. 

We note that for small disk/annulus ($\theta_d \ll 1$ deg), the CMB is remarkably uniform because of the suppression of fluctuation due to Silk damping, and the CIB is always the dominant source of uncertainty. The CMB fluctuations are more important than the CIB at degree scale for $\nu \lesssim 400$ GHz.


\subsection{Equal Width Case}
In this case
\begin{align}
{\rm Disk:} &\ {\rm Radius = \ } \theta_d \\
{\rm Annulus:} &\ {\rm Between \ } \theta_{a,1} = \theta_d {\ \rm and \ } \theta_{a,2} = 3 \theta_d
\end{align}

\begin{figure}
\centering
    \includegraphics[width=8cm]{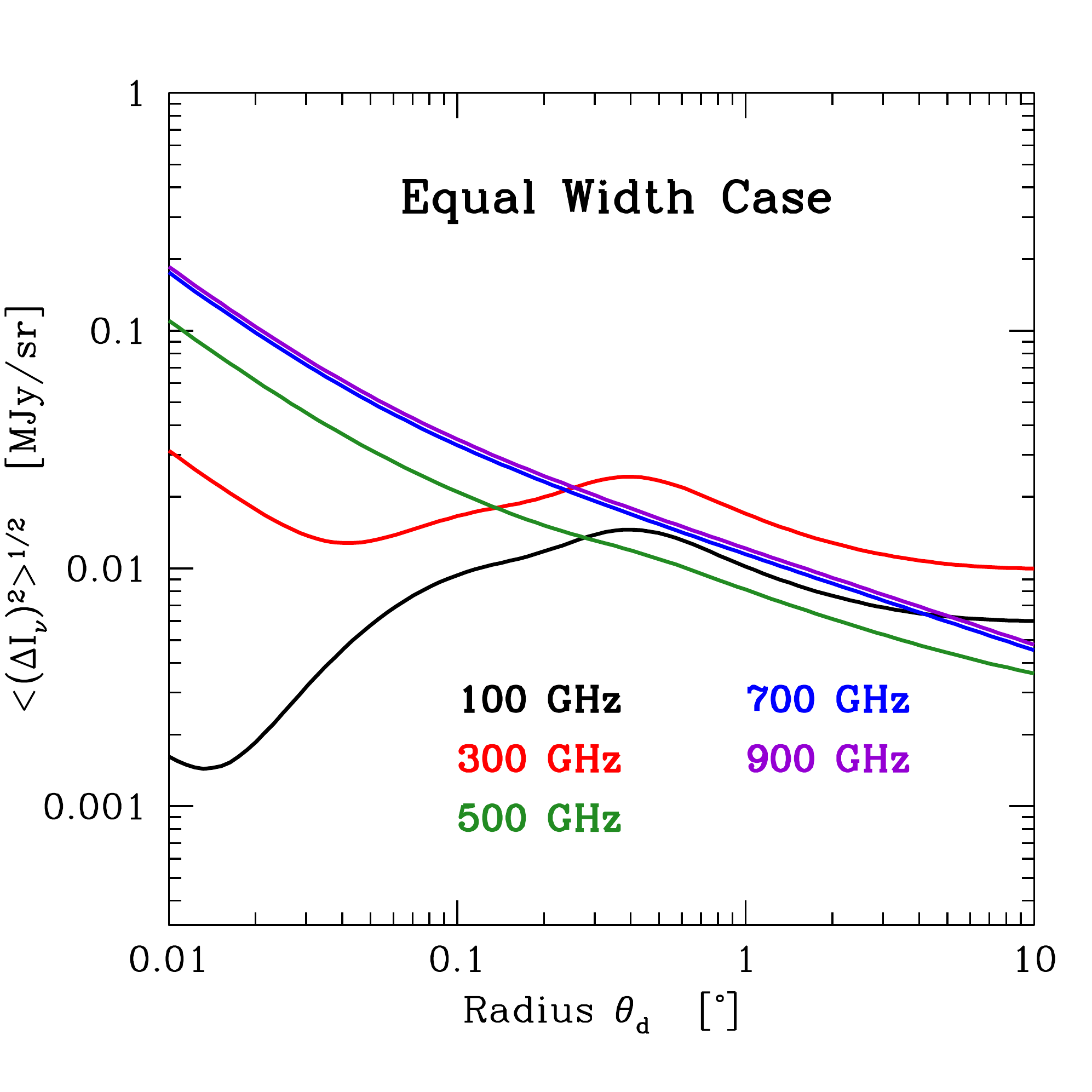}
\caption{Fluctuations in intensity for the equal width case in the limit of an infinitely narrow beam ($b_\ell = 1$). The different curves show fluctuations at 100, 300, 500, 700 and 900 GHz.}\label{fig:EW}
\end{figure}

The results for both temperature and intensity fluctuations are shown in Figure \ref{fig:EW}.
Similar considerations to the previous section apply here, with the CIB dominating for small $\theta_d$, and the CMB dominating at degree-scale for $\nu \lesssim 400$ GHz.


\section{Examples from the Literature}
In this section, we consider cases from the literature in which background subtraction was performed at submillimetre and millimetre wavelengths and illustrate how our formalism would quantify the background subtraction uncertainty in each case.

The SMC is an object of large apparent size ($\sim 4^\circ$) whose submillimetre emission has been a subject of interest in part due to a reported excess relative to predictions by conventional dust models \citep{Israel+etal_2010, Bot+etal_2010, Planck_SMC_2011}. \citet{Israel+etal_2010} report a total intensity of $0.010$ MJy/sr in the 60.7 GHz WMAP band after performing background subtraction with an equal area annulus. \citet{Bot+etal_2010} find that this value is in excess of predictions by dust models by $0.0075 \pm 0.0011$ MJy/sr . Using the formalism we have presented, background subtraction using a $2^\circ$ radius aperture with $\theta_{FWHM} = 21'$ \citep{Israel+etal_2010} and an equal area annulus yields an expected uncertainty of 0.0028 MJy/sr due to background fluctuations. Thus, we find that the reported 60.7 GHz excess of $0.0075$ MJy/sr is unlikely due entirely to background fluctuations. \citet{Bot+etal_2010} explicitly consider the possibility of CMB fluctuations accounting for the excess and perform an annulus subtraction on Monte Carlo realisations of the CMB to quantify the effect. They present a histogram of what we term $\mathcal{D}_{d,a}$ having $\sigma \simeq 0.0025$ MJy/sr, in agreement with the result of our calculation. We note that neglecting contributions from the CIB and radio point sources introduces minimal error at this frequency and aperture size, but stress that at higher frequencies possible CIB contributions should be explicitly considered.

\citet{Planck_SMC_2011} present a re-analysis of the SMC SED using \emph{Planck} observations. They perform background subtraction in two stages. First, following the methods described in \citet{Eriksen+etal_2004}, they take advantage of the different frequency dependence of CMB fluctuations relative to galaxy emission to estimate the CMB temperature over a $2.38^\circ$ radius disk centered on the SMC. They derive an upward CMB temperature fluctuation of 0.0063 MJy/sr at 60.7 GHz, where the RMS CMB fluctuation on a disk of that size is 0.0051 MJy/sr at that frequency. They then use an annular subtraction to estimate the foreground emission. With this estimate for the CMB temperature and the revised estimate of the foreground contribution, the derived total 60.7 GHz emission from the SMC itself was 0.00292 MJy/sr. This is only 30\% of the value estimated by \citet{Israel+etal_2010} and \citet{Bot+etal_2010}, underscoring the potential importance of background fluctuations. We note that if the CMB temperature behind the SMC had been estimated with an equal area annulus instead of using more precise multi-frequency techniques, the resulting subtraction uncertainty of 0.0028 MJy/sr would have been comparable to the {\it total} SMC emission at 60.7 GHz.

In addition to the SMC, excess submillimetre emission has been reported in the low metallicity dwarf galaxies NGC 1705 and Haro 11 \citep{Galametz+etal_2009}. Using LABOCA $870 \mu$m data, \citet{Galametz+etal_2009} find excesses of approximately 30 and 35 mJy above the fiducial model for each galaxy, respectively. Background subtraction at $870 \mu$m was performed by placing small circles outside the 144'' diameter photometric aperture then subtracting the mean. If we assume such a method is comparable to our equal area annulus case, and adopting $\theta_{FWHM} = 18.2''$ for LABOCA as used in \citet{Galametz+etal_2009}, we derive a background subtraction uncertainty at $870 \mu$m of 10.8 mJy. If instead the actual subtraction were more akin to using an annulus with thickness of $\theta_{FWHM}$, the uncertainty increases slightly to 11.4 mJy. These values correspond to approximately 10 per cent of the reported flux density of NGC 1705 and 27 per cent of the reported flux density of Haro 11.

\section{Conclusions}
We have presented simple formulae to quantify the uncertainty in background subtraction for a disk-annulus geometry, allowing for straightforward incorporation in reported error bars. Background subtraction uncertainty at small angular scales ($\theta \ll 1^\circ$) is dominated by the CIB, while the CMB dominates at larger scales for frequencies $\lesssim 400$ GHz. Emission from unresolved radio galaxies can also become important at frequencies $\lesssim 150$ GHz.

We demonstrate our formalism on example cases from the literature, where we find that the background subtraction can be appreciable ($\sim 10$ per cent) relative to the reported galaxy fluxes. Fainter sources would certainly require careful treatment of the background subtraction error.

The methods presented in this paper are tailored to aperture photometry of galaxies from $100 - 1000$GHz and to practical methods of background subtraction used in the literature. We do not consider contributions from spatial structure in Galactic foregrounds, which must be estimated separately. 

We have also assumed that the observations are carried out at a single frequency-- leveraging multi-frequency information allows one to better quantify background contributions, particularly in the case of CMB fluctuations which will differ significantly in spectral shape from the target galaxy. If the error estimates obtained via the techniques developed here are significant relative to the flux of the target object, then it may be necessary to resort to more sophisticated methods.

\section*{Acknowledgements}
We thank Bruce Draine for posing the problem that prompted this work and many comments on the draft. We also thank Emmanuel Schaan, Blake Sherwin, Kendrick Smith and David Spergel for very helpful discussions.

\bibliography{main}

\begin{thebibliography}{}
\makeatletter
\relax
\def\mn@urlcharsother{\let\do\@makeother \do\$\do\&\do\#\do\^\do\_\do\%\do\~}
\def\mn@doi{\begingroup\mn@urlcharsother \@ifnextchar[{\mn@doi@}{\mn@doi@[]}}
\def\mn@doi@[#1]#2{\def\@tempa{#1}\ifx\@tempa\@empty
  \href{http://dx.doi.org/#2}{doi:#2}\else \href{http://dx.doi.org/#2}{#1}\fi
  \endgroup}
\def\mn@eprint#1#2{\mn@eprint@#1:#2::\@nil}
\def\mn@eprint@arXiv#1{\href{http://arxiv.org/abs/#1}{{\tt arXiv:#1}}}
\def\mn@eprint@dblp#1{\href{http://dblp.uni-trier.de/rec/bibtex/#1.xml}{dblp:#1}}
\def\mn@eprint@#1:#2:#3:#4\@nil{\def\@tempa {#1}\def\@tempb {#2}\def\@tempc
  {#3}\ifx \@tempc \@empty \let\@tempc\@tempb \let\@tempb\@tempa \fi \ifx
  \@tempb \@empty \def\@tempb{arXiv}\fi \@ifundefined
  {mn@eprint@\@tempb}{\@tempb:\@tempc}{\expandafter \expandafter \csname
  mn@eprint@\@tempb\endcsname \expandafter{\@tempc}}}

\bibitem[\protect\citeauthoryear{{Addison} et~al.,}{{Addison}
  et~al.}{2012}]{2012ApJ...752..120A}
{Addison} G.~E.,  et~al., 2012, \mn@doi [\apj] {10.1088/0004-637X/752/2/120},
  \href {http://adsabs.harvard.edu/abs/2012ApJ...752..120A} {752, 120}

\bibitem[\protect\citeauthoryear{{Bot}, {Ysard}, {Paradis}, {Bernard},
  {Lagache}, {Israel}  \& {Wall}}{{Bot} et~al.}{2010}]{Bot+etal_2010}
{Bot} C.,  {Ysard} N.,  {Paradis} D.,  {Bernard} J.~P.,  {Lagache} G.,
  {Israel} F.~P.,   {Wall} W.~F.,  2010, \mn@doi [\aap]
  {10.1051/0004-6361/201014986}, \href
  {http://adsabs.harvard.edu/abs/2010A%26A...523A..20B} {523, A20}

\bibitem[\protect\citeauthoryear{{Dunkley} et~al.,}{{Dunkley}
  et~al.}{2013}]{2013JCAP...07..025D}
{Dunkley} J.,  et~al., 2013, \mn@doi [\jcap] {10.1088/1475-7516/2013/07/025},
  \href {http://adsabs.harvard.edu/abs/2013JCAP...07..025D} {7, 25}

\bibitem[\protect\citeauthoryear{{Eriksen}, {Banday}, {G{\'o}rski}  \&
  {Lilje}}{{Eriksen} et~al.}{2004}]{Eriksen+etal_2004}
{Eriksen} H.~K.,  {Banday} A.~J.,  {G{\'o}rski} K.~M.,   {Lilje} P.~B.,  2004,
  \mn@doi [\apj] {10.1086/422807}, \href
  {http://adsabs.harvard.edu/abs/2004ApJ...612..633E} {612, 633}

\bibitem[\protect\citeauthoryear{{Galametz} et~al.,}{{Galametz}
  et~al.}{2009}]{Galametz+etal_2009}
{Galametz} M.,  et~al., 2009, \mn@doi [\aap] {10.1051/0004-6361/200912963},
  \href {http://adsabs.harvard.edu/abs/2009A%26A...508..645G} {508, 645}

\bibitem[\protect\citeauthoryear{{Israel}, {Wall}, {Raban}, {Reach}, {Bot},
  {Oonk}, {Ysard}  \& {Bernard}}{{Israel} et~al.}{2010}]{Israel+etal_2010}
{Israel} F.~P.,  {Wall} W.~F.,  {Raban} D.,  {Reach} W.~T.,  {Bot} C.,  {Oonk}
  J.~B.~R.,  {Ysard} N.,   {Bernard} J.~P.,  2010, \mn@doi [\aap]
  {10.1051/0004-6361/201014073}, \href
  {http://adsabs.harvard.edu/abs/2010A%26A...519A..67I} {519, A67}

\bibitem[\protect\citeauthoryear{{Planck Collaboration} et~al.,}{{Planck
  Collaboration} et~al.}{2011a}]{Planck_SMC_2011}
{Planck Collaboration} et~al., 2011a, \mn@doi [\aap]
  {10.1051/0004-6361/201116473}, \href
  {http://adsabs.harvard.edu/abs/2011A%26A...536A..17P} {536, A17}

\bibitem[\protect\citeauthoryear{{Planck Collaboration} et~al.,}{{Planck
  Collaboration} et~al.}{2011b}]{2011A&A...536A..18P}
{Planck Collaboration} et~al., 2011b, \mn@doi [\aap]
  {10.1051/0004-6361/201116461}, \href
  {http://adsabs.harvard.edu/abs/2011A%26A...536A..18P} {536, A18}

\bibitem[\protect\citeauthoryear{{Planck Collaboration} et~al.,}{{Planck
  Collaboration} et~al.}{2013a}]{2013arXiv1303.5076P}
{Planck Collaboration} et~al., 2013a, \mn@eprint {arXiv} {1303.5076}

\bibitem[\protect\citeauthoryear{{Planck Collaboration} et~al.,}{{Planck
  Collaboration} et~al.}{2013b}]{2013arXiv1309.0382P}
{Planck Collaboration} et~al., 2013b, \mn@eprint {arXiv} {1309.0382}

\bibitem[\protect\citeauthoryear{{Planck Collaboration} et~al.,}{{Planck
  Collaboration} et~al.}{2013c}]{2013arXiv1303.5075P}
{Planck Collaboration} et~al., 2013c, \mn@eprint {arXiv} {1303.5075}

\bibitem[\protect\citeauthoryear{{Reichardt} et~al.,}{{Reichardt}
  et~al.}{2012}]{2012ApJ...755...70R}
{Reichardt} C.~L.,  et~al., 2012, \mn@doi [\apj] {10.1088/0004-637X/755/1/70},
  \href {http://adsabs.harvard.edu/abs/2012ApJ...755...70R} {755, 70}

\makeatother
\end{thebibliography}

\appendix

\section{Finite Beam}
\label{sec:beam}
In the Appendix we derive the formulae quoted in the main text. We use the flat sky approximation throughout, which is excellent whenever the patch of sky considered has angular size less than a few degrees.

First, assume that the sky is observed with a finite resolution beam. Then the observed temperature will be a convolution of the true temperature with the beam function:
\begin{equation}
T^{obs} (\btheta) = \int d^2 \theta' \ T(\btheta') b(\btheta - \btheta')
\label{eq:Tobs}
\end{equation}
Where $b(\btheta)$ is the beam function normalised such that 
\begin{equation}
\int d^2 \theta \ b(\btheta) = 1
\end{equation}
\label{lastpage}
Fourier transforming equation (\ref{eq:Tobs}), we find that $a^{obs}(\bell) = a(\bell) b_{\ell}$, so that the observed $C^{obs}_\ell$ are given by
\begin{equation}
C^{obs}_\ell = b_\ell^2 C_\ell
\end{equation}
Therefore, to take into account the finite resolution of the beam, we simply replace the true $C_\ell$ by $b_\ell^2 C_\ell$.
In the common case of a Gaussian beam,
with width $\sigma_b = \theta_{FWHM} / \sqrt{8 \ln (2)}$, 
\begin{equation}
b(\theta) = \frac{1}{2 \pi \sigma_b^2} \exp \left(-\frac{\theta^2}{2 \sigma_b^2} \right)
\end{equation}
and
\begin{equation}
b_{\ell} = e^{-\ell^2 \sigma_b^2 / 2} = e^{-\ell^2 \theta^2_{FWHM} / (16 \ln(2))}
\end{equation}

\section{Fluctuations in a finite region}
In this Appendix we derive the RMS fluctuations in a masked region. 
For what follows, we will need to compute the real space two-point correlation
\begin{align}
\ \langle \Delta T(\btheta) & \Delta T (\btheta')
\rangle = \int \frac{\mathrm{d}^2\ell}{(2 \pi)^2} \int
\frac{\mathrm{d}^2\ell'}{(2 \pi)^2} \langle a(\bell) a(\bell')
\rangle e^{i(\bell' \cdot \btheta' + \bell \cdot \btheta)} \nonumber
\\
\ &= \int \frac{\mathrm{d}^2\ell}{(2 \pi)^2} \int
\frac{\mathrm{d}^2\ell'}{(2 \pi)^2} (2 \pi)^2 b_\ell^2 C_\ell \delta_D (\bell + \bell')
e^{i(\bell' \cdot \btheta' + \bell \cdot \btheta)}
\nonumber \\
\ &=  \int \frac{\mathrm{d}^2\ell}{(2 \pi)^2} \ b_\ell^2 C_\ell
e^{i \bell \cdot (\btheta - \btheta')}
\end{align}
Let the mask (often called window function) have real space profile $W(\btheta)$, normalised such that 

\begin{equation}
\ \int \mathrm{d}^2\theta W(\btheta) = 1
\end{equation}

Then the \textit{average} temperature fluctuation within the masked region is
\begin{equation}
\ \overline{\Delta T} = \int \mathrm{d}^2\theta
\Delta T (\btheta) W(\btheta)
\end{equation}

%

And its RMS fluctuation from the all-sky mean CMB temperature is given by
\begin{align}
\ \langle \left(\overline{\Delta T}\right)^2 \rangle &= \int
\mathrm{d}^2\theta \int \mathrm{d}^2\theta' \langle \Delta
T(\btheta) \Delta T(\btheta') \rangle
W (\btheta) W (\btheta') \nonumber \\
\ &= \int \frac{\mathrm{d}^2\ell}{(2 \pi)^2}
b_\ell^2 C_\ell \int \mathrm{d}^2\theta' e^{-i \bell \cdot \btheta'}
W (\btheta') \int
\mathrm{d}^2\theta e^{i \bell \cdot \btheta} W (\btheta) \nonumber
\\
\ &= \int \frac{\mathrm{d}^2\ell}{(2 \pi)^2} \;
b_\ell^2 C_\ell | W (\bell) |^2 \nonumber \\
\ &= \int \frac{\mathrm{d}\ell}{2 \pi} \; \ell \ b_\ell^2 C_\ell  | W (\ell) |^2
\label{eq:1rms}
\end{align}
In the last line we have made the further assumption that the window function is circularly symmetric about $\btheta = \bf{0}$, so that $W(\bell)  = W(\ell)$.
Note that here the angle bracket $\langle \cdot \rangle$ denotes \textit{ensemble} averages over realisations of the primordial perturbations (or equivalently averages over positions in the sky), while overbars denote the mean temperature within the window function.

\section{Statistics of the difference}
\label{app:diff}
We now wish to compute the statistics of the difference between the means in the disk and the annulus, characterised by window functions $W_d$ and $W_a$ respectively.\\
If $\mathcal{D}_{d,a} = \overline{\Delta T_d} - \overline{\Delta T_a}$,
\begin{align}
&\  \langle \mathcal{D}_{d,a} \rangle  = 0 \nonumber \\
&\  \langle \mathcal{D}^2_{d,a} \rangle =
\langle \left(\overline{\Delta T_d}\right)^2 \rangle + \langle \left(\overline{\Delta
  T_a}\right)^2 \rangle - 2 \langle \overline{\Delta T_d} \, \overline{\Delta T_a}
\rangle
\label{eq:Drms}
\end{align}
The first two terms in the above can be evaluated using eq (\ref{eq:1rms}) and the appropriate window function.
For the last term, we need to compute the covariance

\begin{align}
\langle \overline{\Delta T_d} \, \overline{\Delta T_a} \rangle &=  \int \frac{\mathrm{d}^2\ell}{(2 \pi)^2} \  
b_\ell^2 C_\ell  W_a^* (\bell) W_d(\bell) \nonumber \\
\ &=  \int \frac{\mathrm{d}\ell}{2\pi} \; \ell \ b_\ell^2 C_\ell W_a (\ell) W_d(\ell)
\label{eq:2rms}
\end{align}
In the last line we have assumed that the window function is circularly symmetric, which is true for our disk and annulus. If this is not the case, then we must evaluate the full 2D integral in the first line.

\section{Window Functions}
\label{sec:win}
For concreteness, we choose top-hat disk and annulus window functions in real space as follows:
For a circular disk of radius $\theta_{d}$, we take
\begin{equation}
 W_d(\btheta) = \left\{
     \begin{array}{lr}
       \displaystyle
       \frac{1}{\pi \theta_{d}^2} & \, \mathrm{for} \, |\btheta| < \theta_{d} \\
       \displaystyle
      0  \, & \mathrm{otherwise}
     \end{array}
   \right.
\end{equation}
which in Fourier space is:
\begin{align}
\ W_d\left(\ell\right) &= \int_0^{\theta_{d}} 
\mathrm{d}^2\theta \ e^{-i \bell \cdot \btheta} \frac{1}{\pi \theta_1^2}
\nonumber \\
\ &= \frac{2}{\ell \theta_d} J_1(\ell \theta_d) 
\end{align}
where $J_1$ is a Bessel function of the first kind.
Similarly, for an annulus defined between radii $\theta_{a, 1}$ and $\theta_{a, 2}$, 
\begin{equation}
 W_a(\btheta) = \left\{
     \begin{array}{lr}
       \displaystyle
       \frac{1}{\pi \left(\theta_{a, 2}^2 - \theta_{a, 1}^2\right)} & \,
       \mathrm{for} \, \theta_{a, 1} < |\btheta| < \theta_{a, 2} \\
       \displaystyle
      0  & \mathrm{otherwise}
     \end{array}
   \right.
\end{equation}
Thus we have that
\begin{align}
\ W_a\left(\ell\right) &= \int_{\theta_{a,1}}^{\theta_{a,2}}
\mathrm{d}^2\theta \ e^{-i \bell \cdot \btheta} \frac{1}{\pi \left(\theta_{a,2}^2 - \theta_{a,1}^2\right)} \nonumber 
 \\
\ &= \frac{2}{\ell  \left(\theta_{a,2}^2 - \theta_{a,1}^2\right)}
\left[\theta_{a,2} J_1\left(\ell \theta_{a,2}\right) - \theta_{a,1} J_1\left(\ell
    \theta_{a,1}\right)\right] 
\end{align}

\section{Uncertainties from CMB only}
\begin{figure}
\centering
    \includegraphics[width=8cm]{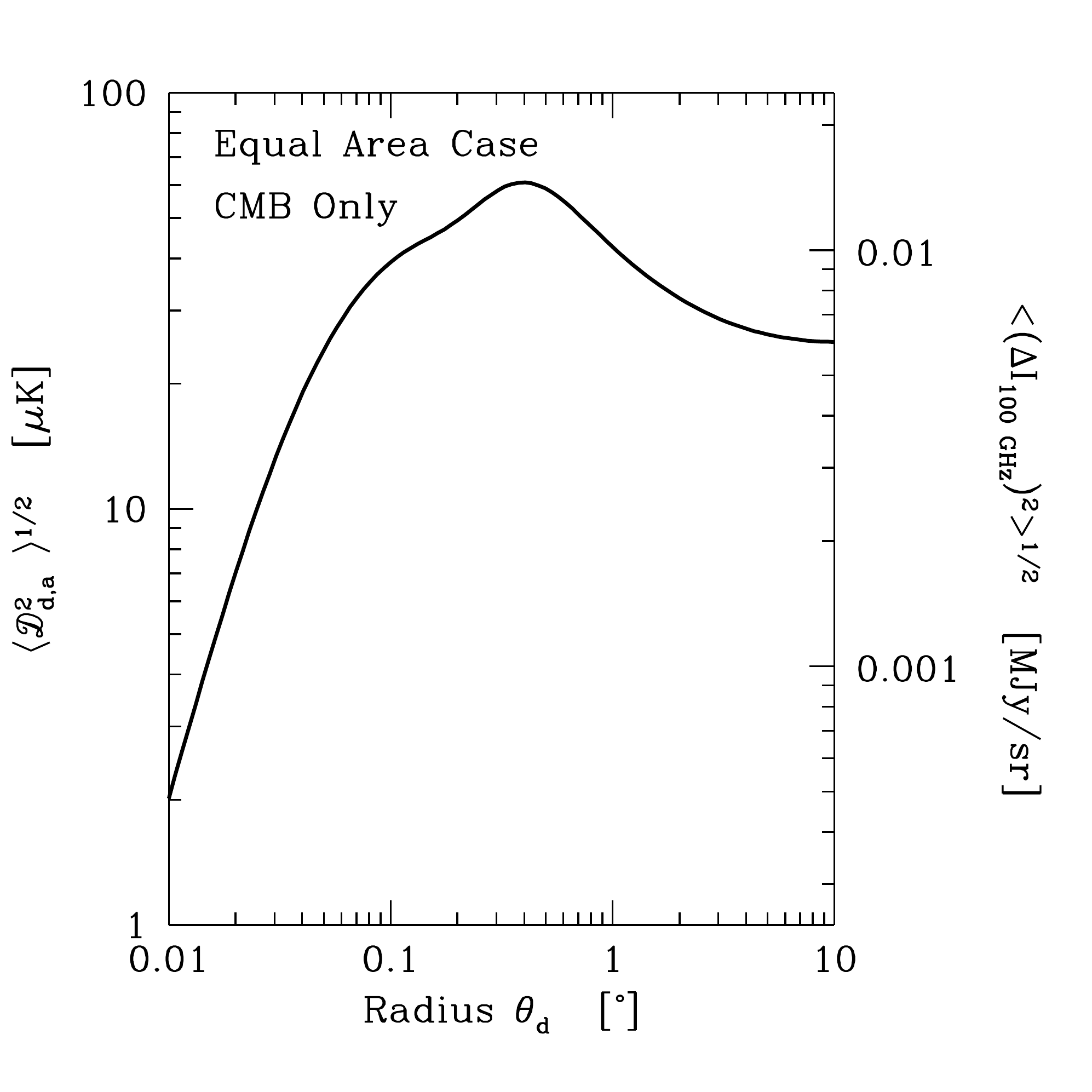}
\caption{Background subtraction uncertainty in the equal area case due to CMB fluctuations only. For convenience, we give the size of the fluctuation in temperature units as well as intensity units at 100 GHz.}\label{fig:cmb_only}
\end{figure}

For convenience we show the uncertainties including the CMB contributions only in Figure~\ref{fig:cmb_only}. 

\section{CIB parameters}
\label{sec:CIBparams}
For the CIB we use parameters from a combined analysis of ACT and SPT data \citep{2013JCAP...07..025D,2012ApJ...755...70R}: $\nu_0  = 150 \ {\rm GHz}, \ \ell_0 = 3000, \ a_p = 7.0 \pm 0.3,\ a_c = 5.7 \pm 0.6,\ \beta = 2.10 \pm 0.07,\ n = 1.2,\ T_d = 9.7$ K. The Planck collaboration finds consistent results, but using Planck data alone, the amplitudes of the Poisson and clustered components are degenerate, due to the lower maximum $\ell$ resolved \citep{2013arXiv1303.5075P}.

\end{document}